# PLOS ONE
# Topic-Level Opinion Influence Model(TOIM): An Investigation Using Tencent Micro-Blogging
## --Manuscript Draft--

| | |
|---|---|
| **Manuscript Number:** | |
| **Article Type:** | Research Article |
| **Full Title:** | Topic-Level Opinion Influence Model(TOIM): An Investigation Using Tencent Micro-Blogging |
| **Short Title:** | Topic-Level Opinion Influence Model |
| **Corresponding Author:** | Daifeng Li, PH.D<br>Tsinghua University<br>Beijing, CHINA |
| **Keywords:** | Tencent Micro-Blogging; Opinion Influence; Sentiment Analysis; Topic Modeling; Indirect Social Influence |
| **Abstract:** | ABSTRACT: Mining user opinion from Micro-Blogging has been extensively studied on the most popular social networking sites such as Twitter and Facebook in the U.S., but few studies have been done on Micro-Blogging websites in other countries (e.g. China). In this paper, we analyze the social opinion influence on Tencent, one of the largest Micro-Blogging websites in China, endeavoring to unveil the behavior patterns of Chinese Micro-Blogging users. This paper proposes a Topic-Level Opinion Influence Model (TOIM) that simultaneously incorporates topic factor and social direct influence in a unified probabilistic framework. Based on TOIM, two topic level opinion influence propagation and aggregation algorithms are developed to consider the indirect influence: CP (Conservative Propagation) and NCP (None Conservative Propagation). Users' historical social interaction records are leveraged by TOIM to construct their progressive opinions and neighbors' opinion influence through a statistical learning process, which can be further utilized to predict users' future opinions on some specific topics. To evaluate and test this proposed model, an experiment was designed and a sub-dataset from Tencent Micro-Blogging was used. The experimental results show that TOIM outperforms baseline methods on predicting users' opinion. The applications of CP and NCP have no significant differences and could significantly improve recall and F1-measure of TOIM. |
| **Order of Authors:** | Daifeng Li, PH.D |
| | Ying Ding |
| | Xin Shuai |
| | Golden Guo-zheng Sun |
| | Jie Tang |
| | Zhipeng Luo |
| | Jingwei Zhang |
| | Guo Zhang |
| **Suggested Reviewers:** | Jiawei Han<br>University of Illinois at Urbana-Champaign, USA<br>hanj@illinois.edu<br>Prof Han is a famous researcher in the domain of social network. |
| | Philip Yu<br>University of Illinois, Chicago, USA<br>psyu@uic.edu<br>Prof Yu is a famous researcher in the domain of social network. |
| **Opposed Reviewers:** | |

**Cover Letter**

Dear Editor:

We are sending a manuscript entitled "Topic-Level Opinion Influence Model (TOIM): An Investigation Using Tencent Micro-Blogging". We would like to submit it to PLOS ONE.

Mining user opinion from Micro-Blogging has been extensively studied on the most popular social networking sites such as Twitter and Facebook in the U.S. The research is important for us to better understand users' behaviors, which could help to provide better online services in the future. But few studies have been done on Micro-Blogging websites in other countries (e.g. China). In this paper, we analyze the social opinion influence on Tencent Micro-Blogging, one of the largest Micro-Blogging websites in China, endeavoring to unveil the behavior patterns of Chinese Micro-Blogging users. This paper proposes a Topic-Level Opinion Influence Model (TOIM) that simultaneously incorporates topic factor and social direct influence in a unified probabilistic framework. Users' historical social interaction records are leveraged by TOIM to construct their progressive opinions and neighbors' opinion influence through a statistical learning process, which can be further utilized to predict users' future opinions on some specific topics.

According to the experiment, users' opinion preferences and influential relationships can be better understood. The model combined with Indirect Influence algorithm shows the dynamic of the structural and semantic features of social networks on Tencent Micro-Blogging to uncover the latent relationships between them. The study contributes for better understanding of the behaviors and relationships of Tencent users, which has a practical value for designing personalized services in Micro-Blogging System and provides a more accurate way to reflect public's opinion in reality.

This paper is a thorough extension of a short paper of CIKM conference (Li, et al., 2012). We have attached the CIKM paper in the attachment. We certify that this paper consists of original, unpublished work which is not under consideration for publication elsewhere.

We also addressed the four questions about the experiment data from Tencent Weibo, which are sent by the editor of PLOS ONE at 26$^{th}$, Sep, 2012. We made the response of four questions as additional materials and attached it behind the manuscript. It is named as "Response To Editors".

We hope that the manuscript meets the high standards of your journal. We are looking forward to receiving a favorable response from you regarding the acceptance of our manuscript.

Sincerely yours.

Daifeng Li (corresponding author)
Address: 1-308 Room, FIT Building, Tsinghua University, Beijing, China
E-mail: ldf3824@yahoo.com.cn
Phone: +86 13611107551
2012-10-09

# Topic-Level Opinion Influence Model (TOIM): An Investigation Using Tencent Micro-Blogging[1]

Daifeng Li[1], Ying Ding[2], Xin Shuai[3], Golden Guo-zheng Sun[4], Jie Tang[1], Zhipeng Luo[5], Jingwei Zhang[6], Guo Zhang[2]

[1]Dept. of Computer Science and Technology, Tsinghua University, Beijing, China

[2]School of Library and Information Science, Indiana University Bloomington, IN, USA

[3]School of Informatics and Computing, Indiana University Bloomington, IN, USA

[4]Tencent Company, Beijing, China

[5]Beijing University of Aeronautics and Astronautics, Beijing, China

[6]Department of Electronic Engineering, Tsinghua University, Beijing, China

ldf3824@yahoo.com.cn, { dingying,xshuai,guozhang}@indiana.edu, gordon.gzsun@gmail.com, jery.tang@gmail.com, patrick.luo2009@gmail.com, iceboal@gmail.com

**ABSTRACT:** Mining user opinion from Micro-Blogging has been extensively studied on the most popular social networking sites such as Twitter and Facebook in the U.S., but few studies have been done on Micro-Blogging websites in other countries (e.g. China). In this paper, we analyze the social opinion influence on Tencent, one of the largest Micro-Blogging websites in China, endeavoring to unveil the behavior patterns of Chinese Micro-Blogging users. This paper proposes a Topic-Level Opinion Influence Model (TOIM) that simultaneously incorporates topic factor and social direct influence in a unified probabilistic framework. Based on TOIM, two topic level opinion influence propagation and aggregation algorithms are developed to consider the indirect influence: CP (Conservative Propagation) and NCP (None Conservative Propagation). Users' historical social interaction records are leveraged by TOIM to construct their progressive opinions and neighbors' opinion influence through a statistical learning process, which can be further utilized to predict users' future opinions on some specific topics. To evaluate and test this proposed model, an experiment was designed and a sub-dataset from Tencent Micro-Blogging was used. The experimental results show that TOIM outperforms baseline methods on predicting users' opinion. The applications of CP and NCP have no significant differences and could significantly improve recall and F1-measure of TOIM.

*Key Words:* Tencent Micro-Blogging, Opinion Influence, Sentiment Analysis, Topic Modeling, Indirect Social Influence

## 1. INTRODUCTION

With the boom of Web2.0, people are inclined to discuss various topics online and express

---

[1] * Corresponding author at: Department of Computer Science and Technology, Tsinghua University, Beijing. Tel: +86 13611107551. E-mail address: ldf3824@yahoo.com.cn (Daifeng Li).

This Paper is a thorough extension of a short paper of CIKM 2012 Conference [1]

their opinions on different topics, ranging from product reviews to political positions. Such a large-scale user-generated text provides a great opportunity for opinion mining, especially in marketing and policymaking. For example, companies can analyze their customers' reviews of their newly released products to figure out their preferences. Governments can supervise public responds to some regulations/policies to gain public support. Recently, Twitter, as the most popular micro-blog site, has arisen as a global domain to spread ideas and opinions all over the world, and has been widely studied [2, 3], but little research has been done to study other micro-blog sites, which are embedded in different cultural, economic and political backgrounds.

With the awareness that cultural and national differences cannot be ignored in studying information transfer, this paper aims to mine opinions from Tencent Micro-Blogging, which is one of the largest micro blogging website in China and the third largest in the world, so as to shed light on users' everyday topics and their attitudes towards various topics. Different from previous studies, this paper endeavors to incorporate topic modeling, social influence and sentiment analysis into one unified model to classify users' polarities, thus to better understand Tencent Micro-Bloggers' behaviors, and to provide an applicable analytical tool which can be widely used in related studies on social media users' behaviors.

Above all, three important principles are identified and constitute the prerequisites for this study: 1) Tencent Micro-Blogging website provides different functions than Twitter to capture enriched communication behavior. 2) Opinions and topics are closely related. 3) Users' opinions are subjective to social influence. Based on these understandings and Tencent Micro-Blogging data, a *Topic-level Opinion Influence Model (TOIM)* is proposed that simultaneously incorporates topic factor and social influence in a unified probabilistic framework. Users' historical messages and social interaction records are leveraged by TOIM to construct their historical opinions and neighbors' opinion influence through a statistical learning process, which can be further utilized to predict users' future opinions towards some specific topics. Besides, TOIM is also valuable for analyzing Tencent data from both microscopic (e.g. A user's opinion preference) and macroscopic (eg. Opinions towards public affairs, Reflections of Social Trends). Based on this model, two topic level opinion influence propagation and aggregation algorithms are developed to consider the indirect influence: CP (Conservative Propagation) and NCP (None Conservative Propagation). We also test and evaluate this model by designing an experiment and using a sub-dataset from Tencent Micro-Blogging.

## 2. RELATED WORK

### 2.1 Sentiment Analysis and Opinion Mining

Online discussions around entities, or objects, often cover a mixture of features/topics related to that entity with different preferential. Several opinion mining related studies are in line with our work. Pang, et al [4] studied the problem of classifying documents by overall sentiment not by topic using machine learning methods. Hu, et al [5] mine opinion features from customers' online reviews. Liu et. al [6] analyzed the sentiment of documents by first extracting the comment target and then to predict the polarity of opinions of the target. Bollen et. al. [2] utilized the public moods mined from Twitter to predict the stock market. Gruhl et. al. [7] predicted book sales by analyzing online chat. Mishne et. al. [8] analyzed Blogger sentiment to predict movie sales. Liu et. al. [9] also used sentiment information from blogs to predict sales performance. However, a common deficiency of all those works is that the proposed approaches extract only the overall sentiment of a document, but can neither distinguish different subtopics within a document, nor analyze the sentiment of a subtopic. Mei et. al. [10] proposed Topic-Sentiment Mixture (TSM) model that can reveal latent topical facets in the combination of a Weblog collection, the subtopics in the results of an ad hoc query, and their associated sentiments. Lin et. al. [11] proposed a joint sentiment/topic (JST) model based on LDA that can detect topic and sentiment simultaneously. Both TSM and JST tried to model topic and sentiment at the same time but social influence is not considered.

Most of existing researches are mainly focused on identifying sentimental polarity of sentences, or detecting a person's opinion from his textual information [12, 6, 13], the main idea of those researches is to first build grammar rules, domain features for sentiment analysis, then apply classification algorithms or network algorithm such as SVM, CRFs or propagation algorithm to learn those rules. The rules are mainly about "*Identification of word's sentiment polarity*"[14, 15, 16], "*Identification of Subjective sentiment and Objective sentiment*"[17, 18, 19], "*Identification of Sentiment Objects*"[3], "*Identification of Opinion holder's attitude towards Sentiment Objects*"[2, 20], "*Sentiment Rules of words combination*"[21], "*Sentiment Rules of the context*"[22] and so forth.

### 2.2 Topic Model based Sentiment Analysis and Opinion Detection

Since the introduction of LDA model [23], various extended LDA models have been proposed for topic extraction from large-scale corpora. For example, Author-Topic (AT) model [24] for detecting authors' topic preferences, Link-PLSA-LDA model [25] for topic based link prediction, Supervised LDA [26] could help to define and generate new topics. In recent years, Topic Model was also applied into the research of sentimental analysis; it is mainly used to detect topics for sentimental analysis. Some researches show that with suitable corpora of sentimental words and prior information, it can gain a high accurate result [11, 27]. Besides, Topic Model could be easily optimized and extended for more complex analysis. In our researches, we combine parse tree and other sentiment analysis algorithms together with User-Link Topic model to realize opinion analysis in social networks, which could obtain a more accurate result for detecting users' opinion preferences on different topics.

### 2.3 Social Influence Analysis and its application in Sentiment Analysis

One main purpose of social influence analysis is to detect and evaluate the existence of social influence [28]. In addition, Kempe et al. [29] constructed a NP-Hard problem to solve influence maximization in social network. Tang et al. [30] learned social influence on different topics and proposed Topical Affinity Propagation (TAP) to model topic-level social influence. Liu et. al [31] designed a LDA based Social Influence model to detect influential relationship among individuals. Crandall et. al [32] developed techniques for identifying and modeling the interactions between social influence and selection by using data from online communities.

In order to make better prediction results, some typical methods are adopted to detect sentiments and opinions from micro blogs. The rise of social media, such as Facebook or Microblog, puts the sentiment analysis in the context of social network. Some existing researches have already considered combining social network with sentiment analysis. For example, Tan et. al. [33] utilized the social connection to improve the sentiment classification performance based on the intuition that connected users are more likely to share similar opinions towards the same entity. Gurrea et. al. [3] applied transfer learning to utilize users' communication features to build opinion agreement graph, thus to infer users' opinion towards each aspects of an event. However, they did not further define a series of topics related to some entities and examine the opinions towards different topics. Furthermore, for the research of influence propagation, Liu et. al. [31] suggested Conservation and Non-conservation for mining indirect influence in different heterogeneous network, but they focused mainly on users' behaviors such as citing, following, and replying.

Those researches provide us a new perspective to investigate social opinion from topic level. We further introduce users' opinions into topic-level social influence to capture users' opinions for different topics in heterogeneous social networks. By simultaneously modeling the social network structure, user behavior, and user opinion preference into a unified model, users' opinions could be predicted to a certain extent, which has strong practical values for social management.

## 3. METHODOLOGY

### 3.1 DataSet

Tencent Micro-Blogging is hosted by Tencent Company, which has more than 370 million users and contributes 30-60 million micro-blogs (a micro-blog can also be defined as a message) each day. It is one of the largest micro blogging website in China and the third largest in the world. Similar to Twitter, Tencent allows users to post messages up to 140 Chinese characters, and users could broadcast their messages, follow others, and all the follower-followee relationships generate a huge social network. But Tencent can also provide many other functions that Twitter lacks. For example, 1) Besides post, repost, mention and reply, Tencent offers novel services such as "comment" and "mention" to facilitate users' interactive communication and self-expression; 2) Tencent provides a multimedia channel in which users can post mood labels, pictures, audios and videos; 3) While the average age of users on Twitter is $39^2$ (35-44 takes up about proportion of 29%), most Tencent users are less than 30 years old (20-29 takes up 60.75%)[3]. The whole Micro-Bloggings dataset was collected from 10/2011 to 1/2012¸ with average 30 million micro-blogs each day, as described in Table 1.

*Insert Table 1 here.*

In Tencent Data, the function of "*Forwards*" and "*Replies*" are same as those in Twitter, while "*Comments*" provides users to make comment for a certain Micro Blog, different with "*Replies*", they are not directly shown on the commenter's micro-blog site but are listed under the original Micro Blogs; In Table 1, the largest number of users' behaviors is "Forwards", the number of "Replies" is small, while "Comments" and are relatively high, which means that users in Tencent usually prefer to express their opinions towards certain Micro Blogs and like to share and discuss it with their friends.

### 3.2 Analytical procedures

As a basic opinion mining task, *sentiment polarity classification* traditionally refers to classification of the polarity of a document into positive or negative. However, in the context of Web2.0, our study shifts from *document-level* to *user-level*, for two reasons. First, user-level sentiment provides different level of opinion than document-level sentiment, especially for behavior-targeted marketing. Second, online generated texts tend to be short and noisy, which makes it difficult and bias to judge the opinions solely from the document itself. Instead, aggregation of texts belonging to one user yields more reliable analysis about that user's opinion.

---

[2] http://royal.pingdom.com/2010/02/16/study-ages-of-social-network-users/

[3] http://data.eguan.cn/yiguanshuju_120638.html

**Opinions and topics** We use a probability value (preferential degree) to describe users' topic preference, and use sentiment polarity (+1, 0, -1) to describe users' opinion or sentiment towards different topics and objects: +1 means opinion agreement or positive sentiment, -1 means opinion disagreement or negative sentiment, and 0 means no opinions or sentiments are detected. Popular topics are naturally discussed heavier than others. Different opinions may be expressed by users towards different topics, where users may like some aspects of an entity but dislike others. For instance, a user may like the picture quality of a camera but dislike its short battery life. If this user favors the picture quality over battery life, he/she may still hold overall positive opinion towards the camera. A voter may support a president candidate's healthcare policy but dissent the foreign policy. If the voter cares more about the foreign policy than healthcare policy, he/she may not vote for the candidate. Both the preferential degree and sentiment polarity varies for different topics related to the same object.

**Users' opinions and social influence** Users not only express their individual opinions, but also exchange opinions with others. Consequently, a user's opinion is affected by, or affects, its neighbors, which can be considered as social opinion influence. *Positive* influence indicates one user share the same opinion with his/her neighbors while *Negative* influence indicates one user holds opposite opinions of its neighbors. Regardless of positive or negative, such influence can be leveraged to infer a user's opinion based on the opinions of the user's neighbors, in addition to the user's individual preference. For example, if two users often talk about different political events, and usually agree with each other, then when a new political event happens, if we know one user's opinion, we could infer that another user may have a high probability to have the same opinion. Besides, "*opinion*" is also propagated through social connections, for example, if user *A*'s opinion is often consistent with user *B* on Topic *Z*, user *B*'s opinion is often consistent with user *C* on Topic *Z*, then user *A*'s opinion will have a high probability to be consistent with user *C* on topic *Z*. Indirect influence within a certain range (2-3 degree) could impact users' behaviors [28]; therefore, it is important to understand the underlying indirect opinion influence in heterogeneous social networks and there lacks of sufficient research at the moment. In this paper, two propagation models, Conservation and Non-conservation are introduced to simulate the opinion indirect influence in Micro-Bloggings.

**Topic level opinion influence analysis** A typical scenario of Tencent Micro-blog is shown in Figure 1. Similar as Twitter, Tencent users can post messages of up to 140 Chinese characters and *follow* other users to read their messages. Two mechanisms are provided to facilitate interaction among users, i.e., *repost* (Which is similar as *retweet* in Twitter) and *reply*. We do not consider the repost interaction in opinion mining because followers merely repeat the original posts without expressing any personal opinions most of the time. By contrast, followers reply to other users' message by leaving their own comment, whose opinions can be mined from the comments. Two types of entities (users and messages) and multiple types of relations (user posts/comments on message, user reply to another user) constitute a heterogenous network built on Tencent Micro-Bloggings.

Specifically, Lucy comments on both Lee and Peggy's messages and replies to both of them on the *visual effect* aspect of the movie *Titanic 3D*. Given the topological and textual information, we generate a topic opinion influence network, with Lucy at the center influenced by Lee and

Peggy. In order to better calculate the influential relationship of *Titanic 3D* among Lucy and Lee, Peggy, their historical communication records should be taken into consideration, first, *Titanic 3D* is mainly about topic movie; second, we abstract all their historical communication records related with movies, and make statistical analysis on how many times they agree with each other, how many times they disagree with each other on movie topic, then a strength value (agree/disagree probability) could be obtained between Lucy and Lee, Lucy and Peggy. If they have common interests preference, then, their value of agreement (agreement could also be seen as positive influence) will be high, otherwise, their value of disagreement (disagreement could also be seen as negative influence) will be low. Finally, if Lee and Peggy have provided their opinions towards movie *Titanic 3D*, then Lucy's opinion towards the same movie could be inferred by jointly considering her own opinion preference and opinion influence from Lee and Peggy.

*Insert Figure 1 here.*

In order to better simulate those processes described above and find the accurate influential relationship among users pairs, the first prerequisite is that whether there exists enough historical communication records for a certain amount of users pairs; the second prerequisite is that whether there exists influential patterns in those communication data sets.

For the first prerequisite, two statistical charts obtained from Tencent Micro-Blogs are summarized in Figure 2, the left one records the number of times each user expressing his/her opinion towards a "Social Security" event, and it satisfies power-law distribution; 4,000+ users express their opinions towards that event for more than 10 times, and around 30,000 users express their opinions more than one times; the phenomenon means that there exists a certain number of users, who tend to continuously express their opinions towards a certain topic. The right subchart records the total number of communication times of two users towards the same "Social Security" event (includes "reply", "mention", "comment"). X-axis is the rank of all users pairs according to their communication times; around 2,000 users pairs chatted with each other about the same event for more than 10 times, and more than 20,0000 user pairs chatted with each other more than 1 time. According to recent studies [33], if two users communicated many times about a certain topic, there may exist some degree of agreement between those two users (e.g. The degree of agreement). Therefore, the communication frequency could help us to better understand the influential relationship among different users.

*Insert Figure 2 here.*

*Insert Figure 3 here.*

Another two examples in Figure 3 are illustrated for explaining the second prerequisite, which is mainly about how users' opinions influence others. Famous users in different domains may cause different opinion influences towards other users. The left sub chart records a famous industry leader's four micro-bloggs and opinions distribution of his repliers at four different time points, the values on blue bar above X-axis represents the percentage of all users who agree with this micro-bloggs, while the values on the red bar under X-axis means disagree; most repliers incline to agree with the opinion of industry leaders; while in the right sub-chart, most users usually do not agree the user (Stock Experts)'s idea at different time points. The reason for that phenomenon is complex, but we could find that there exists some type of influential relationships (e.g. Agree or Disagree) between target users and their followers. By learning those social influences, some potential opinion patterns could be found among individuals.

## 4. MODEL DESCRIPTION

Based on the above investigation, the problem of predicting a user's opinion regarding certain topic can be defined by jointly considering the user's historical opinion and opinion influence from the neighbors. This problem starts with a *Heterogeneous Social Network* on Tencent Micro-Bloggings, where nodes include all users (i.e. followers and followees) and all messages (i.e. posts and comments), and edges include all actions from users to messages (i.e. post and comment) and all connections from users to users (i.e. reply). Specifically, given a query object, a sub-graph $G = (U, M, A, E)$ can be extracted from the Heterogeneous Social Network where $U = \{u_i\}_{i=1}^{V}$ is a set of users once posted or commented on messages about the object, $M = \{m_i\}_{i=1}^{D}$ is a set of messages posted or commented from $u_i \in U$, $A = \{(u_i, m_i) | u_i \in U, m_i \in M\}$ is a set of edges indicating $u_i$ posted or commented on $m_j$, and $E = \{(u_i, u_j) | u_i, u_j \in U\}$ is a set of edges indicating $u_j$ replied to $u_i$. Based on *G*, we list several formal definitions as follows:

*DEFINITION 1. [**Vocabulary**] All distinct words from M constitute a vocabulary $W = \{\omega_i\}_{i=1}^{X}$. According to the word property, we further define noun vocabulary $W_N = \{n_i\}_{1}^{N}$ where $n_i$ is a noun and opinion vocabulary $W_O = \{o\omega_i\}_{i=1}^{A}$ where $o\omega_i$ is an adjective, modal or verb with sentiment polarity. The intuition is that a noun represents a topic while a sentiment word indicates an opinion of the noun.*

*DEFINITION 2. [**Opinion Words**] In a sentence, the opinion about a noun is often expressed by verbs or adjective. E.g. I **like** iphone4, Adele is a **marvelous** singer. Such words are called **opinion words**. We use $O(n_i)$ to denote the opinion word about a noun $n_i$ and $O(n_i) \in W_O$.*

*DEFINITION 3. [**Topic-Noun Distribution**] An object contains several conceptually related topics $T = \{t_i\}_{i=1}^{K}$ and each topic is defined as a multinomial distribution over $W_N$. We define a topic-noun distribution $\Theta = \{\theta_{ij}\}_{K \times N}$ where $\theta_{ij}$ denotes the probability that noun $n_j$ is selected given topic $t_i$.*

*DEFINITION 4. [**User-Topic Distribution**] Users have different preference over a set of topics **T**. We define a user-topic distribution $\Phi = \{\phi_{ij}\}_{V \times K}$ where $\phi_{ij}$ denotes the probability that topic $t_i$ is selected given user $u_i$.*

*DEFINITION 5. [**Topic-User-Opinion Distribution**] Users show different opinions towards the*

same topic. We define a topic-user-opinion distribution $\Psi = \{\psi_{i,j}^k\}_{K \times V \times 2}$ where $\psi_{i,j}^k$ denotes the probability that user $u_i$ prefers opinion $o_j$ given topic $t_k$ and $o_j \in \{-1, 0, +1\}$.

DEFINITION 6. [**Topic Opinion Neighbors**] *For user $u_i$, all users that $u_i$ replied to regarding to topic $t_k$ constitute a set $ON(u_i, t_k)$ which is called $u_i$'s topic opinion neighbors around $t_k$. Each user $u_j \in ON(u_i, t_k)$ can influence $u_i$'s opinion of $t_k$.*

DEFINITION 7. [**Topic-Opinion Influence**] *For any $u_j \in ON(u_i, t_k)$, the influence of $u_j$ on $u_i$ can be measured by $\Omega = \{\omega_{i,j,agree}^k\}_{K \times V \times V \times 2} \cup \{\omega_{i,j,disagree}^k\}_{K \times V \times V \times 2}$ where $\omega_{i,j,agree}^k$ denotes the probability that $u_i$ agrees with $u_j$'s opinion and $\omega_{i,j,disagree}^k$ denotes the probability that $u_i$ disagrees with $u_j$'s opinion on topic $t_k$.*

The most important four parameters are $\Theta, \Phi, \Psi$ and $\Omega$, which bind user, topic, opinion and influence in a unified probabilistic framework. Our task can be reduced to the following two steps:

- First, given *G*, how to estimate $\Theta, \Phi, \Psi$ and $\Omega$?
- Second, given user $u_i$ and topic $t_j$, if $\Theta, \Phi, \Psi$ and $\Omega$ are known, how to predict $u_i$'s opinion of $t_j$ if $u_i$ posts or comments on a new message?

The framework of TOIM is illustrated in Figure 4, which combines the social network, topic modeling and sentiment analysis. The process of $u_1$ posting a message can be modeled as a random process. This user first samples a topic *z* from his/her topic distribution depending on $\Theta$ and then samples noun $n_1$ from the topic-noun distribution associated with *z* depending on $\Phi$. Next $u_1$ finds an opinion word $O(n_1)$ to describe $n_1$ and determine his/her opinion $o_1$ by looking for the polarity of $O(n_1)$. After that, $\Psi$ is updated based according to the value of $o_1$. Another user $u_2$ replies to $u_1$ by commenting on $u_1$'s message. The similar random process applies to $u_2$ and $u_2$ tends to select the same topic *z* as $u_1$. Finally, $\Omega$ is updated based on both of $o_1$ and $o_2$. When $u_2$ posts a new message, $\Psi$ and $\Omega$ can be jointly used to predict $u_2$ opinion of topic *z*.

*Insert Figure 4 here.*

**4.1 Opinion Detection**

Opinion Detection is to capture the opinion word $O(n_i)$ for a noun $n_i$ and judge the polarity of $O(n_i)$ in the context of a message. First, a parse tree[4] developed by FudanNLP group is constructed to exhibit the syntactic structure of a sentence and dependency relations between words. For example, "This product is good and cheap." could be handled by parse tree as below:

*Insert Figure 5 here.*

As can be seen in Figure 5, all words are organized into a tree structure according to their grammar dependency relationship, the string under each word represents parts of speech of each word, for example, VC represents verb, JJ represents adjective, NN represents noun, P represents conjunction, DT represents pronoun. Consequently. $O(n_i)$ can be spotted by analyzing the structure of parse tree. Second, the polarity of $O(n_i)$ is judged by searching a corpus of Chinese sentimental words lexicon provided by Tsinghua NLP group[5], which consists of 5,567 positive and 4,479 negative words. Besides, two additional rules are applied to capture the real sentimental relation: 1) whether there exists negation word, like {*not, don't*}, etc; and 2) whether there exists {*adversative relation*} between $n_i$ and $O(n_i)$, like {*but, however*}, etc.

Based on previous studies, the number of $n_i - O(n_i)$ pairs is usually small (around 5-10% can be detected), due to the short and noisy feature of microblog messages. In order to overcome the limitation of sparse data, we consider the statistical co-occurrence relations from all messages we collected. For each distinct noun $n_i \in W_N$ we find out all adjectives/verbs $o\omega_i \in Wo$ that co-occur with $n_i$ in all messages and pick out the top 20 most frequent co-occurrent $o\omega_1, ..., o\omega_{20}$, which constitutes a set $OS(n_j)$. For each $o\omega_j \in OS(n_j)$, we define a *statistical dependence (SD)*:

$$SD(n_i, o\omega_j) = \frac{CO(n_i, o\omega_j)}{AVEDIS(n_i, o\omega_j)}, j = 1, ..., 20 \qquad (1)$$

Where $CO(n_i, o\omega_j)$ denotes the total number of co-concurrent frequency of $n_i$ and $o\omega_j$, and $AVEDIS(n_i, o\omega_j)$ denotes the average distance of $n_i$ and $o\omega_j$ in all their co-concurrent messages. Then, given a message, if $O(n_i)$ is not found for $n_i$ through parse tree, we can

---

[4] http://code.google.com/p/fudannlp/downloads/list

[5] http://nlp.csai.tsinghua.edu.cn/site2/index.php?option=com_content&view=category&id=36&Itemid=58&lang=en

calculate $SD(n_i, o\omega_j)$ as is shown in Equation 1 and finally obtain a $O(n_i)$:

$$O(n_i) = \underset{o\omega_j \in OS(n_j)}{Arg\ max}\ SD(n_i, o\omega_j) \tag{2}$$

In many cases, a Micro-Blog message may only concern about a single topic, which may conclude several entities, for example, "I support A, but I stand against B", which means that A, B are both about one topic, and their meanings are probably opposite. To handle those situations (detect opposite entities from sentences), a simple competitive vocabulary corpus is generated by applying Topic Models and manual annotation; first, we assign the number of Topics as 50 and run LDA on all nouns from Training Data, then for each topic, we select top 20 ranked words, combine each 2 of them together as candidate set, then we totally get 9,500 entity pairs; at last, we manually label each pair to select entity pairs with opposite meaning, at last, 2,104 entity pairs are found, we name the data set as *CoE*, if *A* and *B* are consistent, *CoE(A, B)*=1, if *A* and *B* are opposite, then *CoE(A, B)*=0.

**4.2 Topic-Level Opinion Distribution**

In order to obtain users' Topic-Level Opinion Distribution, we need to estimate User-Topic $\Theta$ and Topic-Word $\Phi$ distributions firstly, Gibbs Sampling is applied to estimate those parameters, with two prior hyperparameters $\alpha$ and $\beta$, respectively. By applying Gibbs Sampling, we construct a Markov chain that converges to joint posterior distribution on random variables topic *z*, user *x* and noun $\omega$, which can be used to infer $\Theta$ and $\Phi$ [34]. The transition between successive stats of Markov chain results from repeatedly drawing *z* from its distribution conditioned on all other variables. Assuming that $u_i$ posted a message and $u_j$ replied to $u_i$ by adding a comment. If the *l*th noun found in $u_i$'s message is *nh*, we sampled a topic for $u_i$ based on Equation (3).

$$P(z^l = t_k \mid x = u_i, \omega = n_h, Z^{-1}) \propto \frac{C_{xz}^{-l} + \alpha}{\sum_{z \in T} C_{xz}^{-l} + K\alpha} \frac{C_{z\omega}^{-l} + \beta}{\sum_{\omega \in W_N} C_{z\omega}^{-l} + N\beta} \tag{3}$$

Where $z^l = t_k$ denotes the assignment of the *l* th noun in to topic $t_k$ and $Z^{-1}$ denotes all topic assignments not including $n_h$. $C_{xz}^{-l}$ and $C_{z\omega}^{-l}$ denote the number of times topic *z* is assigned to user *x*, and noun $\omega$ is assigned to topic *z* respectively, not including the current assignment for the *l*th noun. For user $u_j$, if $n_h$ also occurs in $u_j$'s replying message, $n_h$ is also assigned to topic $t_k$ and $t_k$ is assigned to user $u_j$. For all other nouns in $u_j$'s replying message, the assignment of words and topics are the executed as the same probability as shown in Equation (3). The final $\Theta$ and $\Phi$ can be estimated by:

$$\theta_{xz} = \frac{C_{xz} + \alpha}{\sum_{z \in T} C_{xz} + K\alpha},\ \phi_{z\omega} = \frac{C_{z\omega} + \alpha}{\sum_{\omega \in W_N} C_{z\omega} + N\beta} \tag{4}$$

Topic-level opinion regarding a topic can be easily obtained through aggregation of all message-level opinion records. We define two counters $C_{i,+1}^{k}$ and $C_{i,-1}^{k}$, $i=1,...,V$, $k=1,...,K$ to record the number of times that user $u_i$ expresses positive or negative opinions towards topic $t_k$ by scanning all $u_i$'s message. Then $\Psi$ can be estimated as:

$$\psi_{i,+1}^{k} = \frac{C_{i,+1}^{k}}{C_{i,+1}^{k} + C_{i,-1}^{k}}, \psi_{i,-1}^{k} = \frac{C_{i,-1}^{k}}{C_{i,+1}^{k} + C_{i,-1}^{k}} \qquad (5)$$

**4.3 Topic-Level Opinion Influence**

In addition, we define another two counters $C_{i,j,agree}^{k}$ and $C_{i,j,disagree}^{k}$ to record the number of times $u_i$ and $u_j$ agree or disagree on topic $k$ by scanning all their "post-reply" messages. In our research, "agree" and "disagree" for a certain topic $k$ includes three situations, first, two users hold the same opinion towards the same entity related with topic $k$; second, two users hold same opinion towards different entities A and B, which are related with topic $k$, if A and B have opposite meaning, then two users have "disagree" on topic $k$, else if A and B have consistent meaning, then two users have "agree" on topic k; third, two users hold different opinions towards different entities A and B, which are related with topic $k$, if A and B have opposite meaning, the two users have "agree" on topic $k$, else if A and B have consistent meaning, the two users have "disagree" on topic $k$. Then $\Omega$ can be estimated as:

$$\omega_{i,j,agree}^{k} = \frac{C_{i,j,agree}^{k}}{C_{i,j,agree}^{k} + C_{i,j,disagree}^{k}}$$

$$\omega_{i,j,disagree}^{k} = \frac{C_{i,j,disagree}^{k}}{C_{i,j,agree}^{k} + C_{i,j,disagree}^{k}} \qquad (6)$$

The strength of tie is also important to determine the opinion influence from neighbors, regardless of positive or negative influence. Strength of tie is defined as: for a target user X, other users' influential strength towards him; strength of tie can also be considered as confidence of opinion influence relationship, for example: if two users have agreed with each other for only one time, then the confidence of their influential relationship is relative low, otherwise, the confidence of two users' influential relationships is high. Especially, for $u_i \in ON(u_j, t_k)$, we calculate the strength of relation by:

$$s_{i,j,agree}^{k} = \frac{C_{i,j,agree}^{k}}{\sum_{u_i \in ON(u_j, t_k)} C_{i,j,agree}^{k}}$$

$$s_{i,j,disagree}^{k} = \frac{C_{i,j,disagree}^{k}}{\sum_{u_i \in ON(u_j, t_k)} C_{i,j,disagree}^{k}} \qquad (7)$$

In many cases, given a pair $u_i$ and $u_j$, we could detect both of their opinions towards two entities respectively, but we could not judge whether their attitudes towards a topic is "agree" or "disagree", for example, unidentified relationships between two entities. To solve the problem, we need to utilize other information in addition to the content of their messages. According to previous studies [3, 33], A metric *Opinion Agreement Index (OAI)* is introduced to quantify the influence of $u_i$ on $u_j$:

$$OAI(u_i, u_j) = a \times Influence(u_i) + b \times Tightness(u_i, u_j) + c \times Similarity(u_i, u_j) \qquad (8)$$

Where $Influence(u_i)$ is a function of the number of $u_i$'s followers, we first rank all $u_i$ according to their number of followers, assume the rank of $u_i$ is $Rnk\_Followers(u_i)$, then $Influence(u_i) = \dfrac{1}{Rnk\_Followers(u_i)^\lambda}$; $Tightness(u_i, u_j)$ is the function of the frequency of interactions (we mainly consider "Reply", "Comment", "Mention" ) between $u_i$ and $u_j$, assume $Rnk\_Interactions(u_i, u_j)$ is the rank of different users pairs' communication times, then $Tightness(u_i, u_j) == \dfrac{1}{Rnk\_Interactions(u_i, u_j)^\lambda}$; $Similarity(u_i, u_j)$ is the cosine similarity between $\theta_i$ and $\theta_j$. $a$, $b$, $c$ and $\lambda$ are assigned as 0.6, 0.3 and 0.1 based on empirical knowledge, respectively. $OAI(u_i, u_j)$ is generally normalized for $u_j$:

$$NOAI(u_i, u_j) = \dfrac{OAI(u_i, u_j)}{\sum_{u_i \in ON(u_j, t_k)} OAI(u_i, u_j)} \qquad (9)$$

If $u_j$ replies to $u_i$'s one message and either or none of their opinions can be determined, then $NOAI(u_i, u_j)$ can be used to approximate the probability that $u_j$ agrees with $u_i$ in this replying behavior.

**4.4 Opinion Influence Propagation**

The introduced model in above sections only discovers direct opinion influence, but does not consider indirect influence. In reality, indirect influence could also affect users' behaviors [31, 33], the influence mechanism can be described as below:

*Insert Figure 6 here.*

As can be seen in Figure 6, there is no direct opinion influence between U1 and Ut, while for indirect influence, U1 has direct influences on U2, U3,… Uk; U2, U3, …Uk have influences on Ut; then U1 might indirectly influence Ut; for example, if we know that Uk disagrees with U1 on an entity *o*, and Ut agrees with Uk on the same *o*; then we could infer that Ut may not agree with

U1 on *o*. Except for building influential relationship between two users, indirect influence could also be used to refine the relationship between two users, for example, if two users only communicated with each other for one time, then the calculated relationship may be not correct, if we consider the factors of indirect influence, a more accurate relationship between the two users may be inferred. In order to compute indirect influence, two types of diffusion process conservative and non-conservative are introduced [31] to obtain indirect influence strength. The definitions of conservative and non-conservative are summarized below:

*DEFINITION 8* Influence Propagation: For a propagation process in a graph *G* with *N* nodes, *S* is a *N* dimension vector representing the influential weight of each node. *F(S)* is a mapping function to calculate the new influential weight of each node after Influence Propagation.

As can be learned from formula (5), the influence strength $S_{U_i}^Z$ of user $U_i$ for topic Z in a graph *G* can be defined as:

$$S_{U_i}^Z = \sum_{U_j \in N, U_i \neq U_j} S_{U_i, U_j, agree}^Z \tag{10}$$

Then *S* under a certain topic Z can be defined as $S=\{ S_{U_1}^Z, S_{U_2}^Z, S_{U_3}^Z, ... S_{U_N}^Z \}$.

*DEFINITION 9* Conservative Propagation: For a propagation process in a graph *G* with *N* nodes, if $\| S \|_1 = \| F(S) \|_1$, then we define the process as Conservative Propagation.

*DEFINITION 10* Non-Conservative Propagation: For a propagation process in a graph *G* with *N* nodes, if $\| S \|_1 \neq \| F(S) \|_1$ then we define the process as Non-Conservative Propagation.

We implement both Conservative and Non-Conservative Propagation into our model [31] to re-calculate the influence weight between different nodes, we define $(\Delta S_v^Z)^0 = \{ S_{v,U_1,agree}^Z, S_{v,U_2,agree}^Z, ... S_{v,U_N,agree}^Z \}$, then for Conservative Propagation, the formula can be seen as below:

$$F_t(\Delta S_v^Z) = (1-\beta) \times \sum_{i=0}^{t-1} (\beta^i \times (\Delta S_v^Z)^i) + \beta^t \times (\Delta S_v^Z)^t \tag{11}$$

Where $(\Delta S_v^Z)^i = (\Delta S_v^Z)^0 \times TM^i$, *TM* is an adjacency matrix, $TM(i,j) = S_{U_i,U_j,agree}^Z$.

Following the same way, for Non-Conservative Propagation, the formula can be seen as below:

$$F_t(\Delta S_v^Z) = \sum_{i=0}^{t} \beta^i \times (\Delta S_v^Z)^i \tag{12}$$

Both Conservative and Non-Conservative Propagation could be applied to re-distribute the weight of different nodes based on indirect influence.

**4.5 Opinion Prediction**

Our ultimate goal is to predict a user's opinion about a topic given his/her own opinion preference and his/her neighbor's opinion. First, we need to estimate four parameters $\Theta$, $\Phi$, $\Psi$ and $\Omega$. A Gibbs-Sampling based parameter estimation algorithm is proposed, where topic modeling, sentiment analysis and influence analysis are interwoven together, shown in Algorithm

1. Note that **Pre** 1 to **Pre** 5 should be executed before entering the loop. In each iteration, Gibbs sampling is used to assign nouns to topics and topics to users, and parse tree and *NOAI* are used to detect the opinion polarity. When the iteration is done, the four parameters are calculated.

*Insert Algorithm 1 here.*

Based on the learning results, we would like to predict users' opinion towards some objects with different topic distributions (eg, a new movie, the trend of stock price, a famous person). Two factors are taken into consideration for opinion prediction. First, the historical records of topic preference and opinion distribution learned from TOIM; second, the historical opinions of neighbors and their influence types and strengths learned from TOIM. The prediction result is sampled from a sum-of-weighted probability combing the two factors together as a random process, which is illustrated in Algorithm 2.

*Insert Algorithm 2 here.*

**4.6 Distributed TOIM Learning**

As QQ Micro-Blogging contains millions of users and hundreds of millions of social ties, it is impractical to learn TOIM from a huge dataset using a single machine. Big data may cause two problems: memory space and computing time. Data separating technology and sparse representation could be applied to solve the first problem. To speed up computing, we deploy the learning task based on the Map-Reduce Framework.

First, **Pre**-1 to 3 in Algorithm 1 could be handled by applying simple Map-Reduce task; second, for computing the influential relationship, a Map-Reduce based algorithm is designed for detecting opinions of "*x-xr*" user pairs. We separate original algorithm into two parts, the first part is to run a user-topic model individually to get matrixes of user-topic distributions and topic-word distributions; the second part is to compute the opinion influential relationship for all user pairs based on results from the first part, we speed up the second part using the Map-Reduce process. In Map process, we distribute all "*x-xr*" pairs as Key, their micro-blogs "submit-reply" pairs as Value. In Reduce process, all the same "*x-xr*" pairs are clustered into the same iteration and different iterations are distributed into different machines. For different iterations, similar method as algorithm 1 could be applied to capture users' opinion and thus get influential relationship for different "*x-xr*" user pairs. Different with basic TOIM, it is very fast that all related content of "*x-xr*" are collected into one iteration, so counter variables are only needed to design for each "*x-xr*" pair. The output of each "x-xr" pair is in text format, which is <*x-xr*, {*i: j: k: s_value*}>, where *i* is the number of topic, *j* is the opinion of *x*, *k* is the opinion of *xr*, and *s_value* is the influence weight with fixed *i, j, k*. The process of calculating the influential relationship of "*x-xr*" pair is defined as "*Pair_Computation*".

Users' opinion influential relationship could be detected by the proposed model based on their historical communication records, then opinion prediction can be made by applying the learned model. The function of inferring users' opinion is valuable for marketing strategy, public affairs and so forth.

## 5. RESULTS AND DISCUSSION

In order to validate and testify this proposed model in real data, an experiment is designed and a sub-dataset from Tencent Micro-Blogging is used, to use map-reduce framework for scaling up the efficiency, to apply this model to make opinion prediction and analyze opinion influence on both *microscopic* and *macroscopic* levels, with the endeavor to face challenges, improve our model and to inspire future works.

**5.1 Sub-dataset and sampling**

The whole data set is collected from Tencent QQ micro-blogs that were posted between October 1, 2011 and January 5, 2012. To better test the proposed model, firstly 5 hot objective keywords (which are discussed highly in Tencent Micro-Bloggins) during those 3 months are selected as experiment objects: "$O_1$-Muammar Gaddafi; $O_2$-The Flowers of War (a new Chinese movie); $O_3$-Chinese economics; $O_4$-School bus accident; $O_5$-a college talk from the president of Peking University". Each object contains 10 million level micro blogs. Then the 5 hot objects are used as keywords to retrieve all the related messages and users. The main purpose of this retrieval is to predict users' behavioral patterns from their history interaction records.

For all retrieved users, the Top 1,000 most active ones are selected and their published micro-blogs (include "*initiator, comment, reply, mention*") during those 3 months are searched. Specifically, for user X's "*comment, reply and mention*" behaviors, the information of his parent node user XZ (i.e. who X is currently communicating with) and of root node user XR (i.e. who provides the original discussed topic or message for X), is crawled. Then all historical communication records of X and XZ, as well of X and XR, are crawled. Based on these results, users' social networks for each object (or keyword) are built up, which is described in Table 2.

*Insert Table 2 here.*

For each object, all messages are ordered chronologically and the last 200 (messages should have significant opinion preference, such as "I support this man.") are selected as the testing sample, which includes a total number of 1,000 messages. The rest messages are used for training purpose.

**5.2 Testing procedures**

**Comparison testing** Three algorithms -- SVM (Support Vector Machine), CRF (Conditional Random Field), JST (Joint Sentiment Topic)—are used to make comparison. The traditional ideas to predict users' opinion toward a certain object are based on their historical behavior records, but ignoring the influential relationship, especially the indirect influence, among users. As classical classification algorithms, SVM and CRF are widely applied in sentiment analysis [12,14,33,35]. JST is a probability model to estimate users' opinion preference on different topics. Different with our proposed algorithms, the three algorithms do not take the influential relationship among users into account.

To generate the training dataset, four attributes of each user are defined as: 1) username, 2) nouns with their weighted score, 3) qualifiers of the nouns and 4) topics ID related with those nouns, which are the same for all three algorithms. For example, a user $X_t$ writes a micro-blog *mb*, then the input format should be: *{Opinion (+1, 0, -1); Username:Weight0; Noun1:Weight1; Qualifier1:Qweight1; Noun2:Weight2; Qualifier2:Qweight2;...Topic ID:WeightZ}*. *Noun$_i$* (i=1,2…) is a word, which is mentioned in *mb*. KeyWord Extraction technology[6] is used to compute the weight of each noun and their qualifiers according to their grammar position in *mb*: 1). The noun with highest score should be the most important core words. 2) User's attitude toward this noun could be considered as the input of "*Opinion*". 3) "*Weight0*" of "*Username*" is assigned as 1, while "*WeightZ*" of "*Topic ID*" is the score of "*Topic ID*" on *mb*. Thus, all three algorithms use users' associated attributes and their auto-labeled opinions for each of their micro-blogs to train a classification model and then apply the trained model to predict users' opinion. For SVM, we adopt SVM-light[7]; for CRF, we adopt the code provided by Tang et. al[30] in 2011; for JST, we develop an algorithm to realize its function based on Lin's work [11].

---
[6] http://code.google.com/p/fudannlp/

**Evaluation measures**    Three measurements are used to evaluate the methods:

• **Prediction.** We evaluate the proposed model in terms of Precision, Recall, and F1-Measure, and compare with the baseline methods to evaluate the proposed model.

• **Efficiency.** We use the run time to evaluate the parallel processing of the algorithm.

• **Case studies**. We use several case studies as the anecdotal evidence to further demonstrate the effectiveness of our method.

The basic learning algorithm is implemented in C++ and all experiments are performed on a servers cluster with 36 machines, each of which contains 15 Intel(R) Xeon(R) processors (2.13GHZ) and 60G memory. The data set is stored in HDFS (Hadoop Distributed File System).

### 5.3 Testing results

**Prediction Performance**

We first use all the training data to train TOIM model and use the learned model to predict users' opinions for different objects. Figure 7 shows the different evaluation results for assigning different number of topics.

As can be seen in Figure 7, precision of TOIM has a positive correlation with the number of Topics, when the number of topics is small, the positive correlation is more significant; when the number of topics is big enough, the precision will not make more improvement and begin to fluctuate in a certain range. The highest precision could reach up to 75%, which means the proposed algorithm could successfully detect potential users' behaviors from large-scale micro-blogs dataset. While the performance of SVM and CRF seems to gain relative low scores, which do not be significantly influenced by the number of topics. For JST, it's more likely a random process without the prior knowledge from his neighbors. Besides, the recall of TOIM is relative low (Range is from 0.175 to 0.324), compared with other algorithms (e.g., SVM is from 0.641 to 0.861). The reason is that TOIM cannot successfully detect the opinion or sentiment of many micro-blogs due to the limitation of grammar analysis technology, thus it gains a low recall. The reason for SVM, CRF to have a high Recall is that, those algorithms do not mainly consider the grammar factors but focus more on mining users' opinion preferences based on the provided value of attributes, which are abstracted from each micro-blog and usually can not represent the true meaning of that micro-blog.

*Insert Figure 7 here.*

Next, we apply Conservative and Non-Conservative algorithms to re-calculate the opinion influence weight of different users, and re-calculate weights to make predictions for the same dataset. According to previous studies [31, 33], which are mainly about social influence, we assign the number of *t* as 1, 2 and 3 respectively, and compare each result with TOIM (which did not consider the indirect influence), the comparison results and significant test are summarized as below:

---

[7] http://svmlight.joachims.org/

*Insert Table 3 here.*

As can be seen in Table 3, though it seems there exists positive performance improvement for Conservative and negative performance improvement for Non-Conservative, *t-test* for both propagation is not significant, so the performance of Conservative and Non-Conservative for precision are limit. While for recall, both Conservative and Non-Conservative gain a better performance than TOIM, thus significantly improve the value of F1-measure, the result can be seen in Figure 8 and Figure 9.

As Figure 8 shows, the average Recall of TOIM & Indirect Influence means that we calculate the average value of recall from both Conservative and Non-Conservative (we assign *t=2*, because according to our experiments, Recall seems not significantly influenced by the value of *t*), the value of average recall is significantly improved compared with TOIM. The reason is that indirect influence could help to build influence connection between two users, who are without direct connections, which could help to improve the prediction of users' opinion.

*Insert Figure 8 here.*
*Insert Figure 9 here.*

We further test F1-Measure on dataset $O_1$, $O_2$ and $O_3$ of TOIM and TOIM with Indirect influence; because of the limitation of sentiment detection technology, which makes the train data very sparse, F1-measure of Recall is not very good (If we want to obtain a high precision, then the identification of opinions should be more strict, thus the opinions of many micro blogs could not be detected), however, we find that integrating TOIM with Indirect influence algorithms could significantly improve recall and F1-Measures. The reason is that Indirect Propagations could help to provide potential influential users for opinion predicting, for example, if we want to predict a user's opinion toward a certain Entity *E*, in many cases, we could not find his/her direct neighbors, who have expressed their opinions towards *E*, but if we take all 2 or 3 steps indirect influential users into consideration, then we will gain a higher probability to predict users' behaviors, due to the limitation of opinion detection algorithm, the precision does not improve significantly, but the recall and F1-Measure will be improved significantly.

**Efficiency Performance**

Dataset from Object_1 (which contains 320,176 micro-blogs) is used for efficiency test. As can be seen in Figure 9, the left figure illustrates the time cost of basic TOIM and Map-Reduce based Distributed TOIM (We only make Map-Reduce for "*Pair-Computation*" process, which is introduced in Section 3.5) with different number of Topics, basic TOIM needs average 40 minutes to handle 320,176 micro-blogs (the number of iteration is set as 50), while by applying "*Pair-Computation*" (not considering Topic running process which takes about 5-30 minutes on a single machine with 50 iterations), total average time for detecting users' opinion and influential relationship patterns is about 190 seconds. So the total time of Distributed TOIM is around 33 mins. While in the right figure, the performance of Distributed TOIM is similar with basic TOIM. We also use Distributed TOIM to make experiments on other Object dataset, the results show that the distributed strategy could easily handle millions level of data and significantly speed up the learning process.

*Insert Figure 10 here.*

**Qualitative Case Study**

By applying TOIM, we mainly analyze the algorithms from three levels: user level, local influence level and global influence level to demonstrate the function of our proposed model.

**Users Level:** One of the important output of TOIM is that it could detect single users' opinion preference on different topics. Some representative results are listed in Table 4 below:

*Insert Table 4 here.*

By applying "Identify Certification" Service, which is provided by Tencent, some famous people and their interesting preferences could be detected from TOIM (Due to the Privacy Protection mechanism, many other accounts can not be detected). Four representative users and their favorable topics are chosen. Although the four users's average number of messages is around 50, they are still detected as the most influential persons in their topics, because according to TOIM, if a user's opinions are popular and discussed by many others, he will gain a high score of Influence. The third column lists the frequent discussed topics of each user, the words in bracket are the expression of that topic, the value behind each Topic is the weight of users' opinion preference; for example, for Topic 1, user 622007070 will have a probability of 0.7101 to pick positive opinion, and 0.2899 to pick negative opinion. The words in the fourth column behind each topic means the most representative word, which has be often used to describe current user's opinion on certain topic; for example, "*University*" means that user 622007070 have ever held positive opinions toward that word on topic 1. Above all, by observing results from users level, we could detect users' concentration on what topic, and their attitudes toward different objects. Furthermore, Some behaviors patterns could also be detected to help to analyze users' personalized features, for example, 622007070 and 394865678 seems to hold a relatively moderate attitudes towards Topic 1, and near-neutral attitudes towards other Topics. While user 1484189007 and 790147482 seem more aggressive, and often brought forward penetrating opinions towards Topic 13, 14, 27.

***Microscopic* Influence Level:** *Microscopic* influence level analyzes how a user's opinion influences others on a certain topic. By applying TOIM, direct influence could be obtained among users, an intuitive example of direct influence could be seen in Figure 11:

*Insert Figure 11 here.*

As can be seen in Figure 11, a famous economic researcher, "*Xianping Lang*" is selected, which is in the center node with his figures, the direct lines from center node to others means that he has an positive opinion influence on those users on Topic *Economics*, nodes with figures are the detected famous persons (the followers of each one is bigger than 100 thousands), the values on each line means the influential degree, for example, for node 815902964, the value 0.53 means that when "Xianping Lang" expresses a positive opinion towards Topic Economics, the node will have 0.53 probability to agree with him. The values in the right box record number of "*agreement*" times of "Xianping Lang" with other users, for example, "**UID-**1516940868: 72(Times of Agree)" means that according to their communication records, user 1516940868 has 72 times to agree with "Xianping Lang"'s opinion in economics domain. Those values could be considered as a confidence for influential degree, if two users have agreed with each other for many times, then we have more historical record to learn and the influential degree between them is more accurate.

Another example is to observe the indirect influence of users by Combining TOIM with Conservative Propagation algorithm (it seems to gain a better result than Non-Conservative Propagation). By applying Conservative Propagation, a user's influential degree could also be re-calculated, similar with PageRank, if he/she has influence on a more influential persons, then his/her influential degree will gain a significant improvement. Figure 12 provides us an intuitive

example to illustrate that phenomenon:

*Insert Figure 12 here.*

In Figure 12, the big red node represents the target users with ID as 1032876641, he has positive influential relationship with all other users, which are represented as red nodes in this Figure. For each connection line, the arrow direction from node X to node Y means X has a positive influence on Y in current Topic: "*College & Education*", the number on each line means how many times two nodes have made opinion agreement on current Topic, which can be calculated from Equation (6). Node 1032876641 has direct influence on a small number of nodes (31.2%), while he has indirect influence on all other nodes (68.8%). Through user registration information, we can find out that the UID is a student, he only has 1,780 followers, but he gains a very high rank in College & Education domain, the reasons is that he likes to communicate with many high influential persons, institutions (We only list three here with their figures and text box of descriptions) by using replies, some influential users also gave feedback to his opinions, which significantly improve this user's influential degree calculated by Conservative Propagation.

The following example will illustrate how TOIM improves the performance of opinion prediction by applying social networks. We take "*Muammar Gaddafi*" as evaluation target, and 5 representative users are selected. We use TOIM and CRFs to predict their opinions towards "*Muammar Gaddafi*", the results could be seen as below:

*Insert Table 5 here.*

As can be seen in Table 5, five instances are selected to show that TOIM could use the attributes of social networks to make better predictions. For the five users listed in the first column, it is hard to detect their opinions towards "*Muammar Gaddafi*" only by analyzing their history records by using CRFs (as can be seen in the last column). But if we consider social networks into the model, and could find their opinion influential relationships, then the performance will be improved by considering the known opinions of their neighbors. For example, if we want to infer user Xgdd's opinion towards "*Muammar Gaddafi*", then we can use her neighbors' opinions to make judgment, six high influential neighbors and their opinions are summarized for Xgdd in Table 5, "positive" means that users hold positive opinions towards "Muammar Gaddafi", while "Negative" means users' attitudes are negative; "Agree" means according to historical communication records, Xgdd often agrees with those users' opinions, while "Disagree" means Xgdd did not agree with them. So if we could get Xgdd's neighbors' opinions, then Xgdd's opinion could be inferred by algorithm 2. The results show that, for some users, if there exists sufficient information of their neighbors, TOIM could detect their opinions more accurately.

*Macroscopic* **Influence Level:** First, by applying TOIM and Conservative Propagation, we could quickly detect the most influential users for different topics. We selected 5 popular topics, which are mainly about Topic A: College & Education, Topic B: Daily Emotion, Topic C: Chinese Economics, Topic D: Economics & Tech, Topic E: International Political. For each topic, after removing account of online service institutions (those institutions, such as news, jokes, fashions and et. al often gain high attentions and users tend to express their opinions towards their micro-blogs), we find that many famous persons can be detected at the top 20 influential users for each topic, and some of their information are summarized below in Figures 13:

*Insert Figure 13 here.*

As can be seen in Figure 13, many famous persons could be detected by using TOIM and

Conservative algorithm, which means that those users' opinions towards a certain topic are often supported by many other users, all of them have a large amount of followers, the related topics are also consistent with their backgrounds, for example, user 622002188 is a famous expert in education domain, user 622004678 is a famous economist, while user 1516940868 is a famous international journalist, who is professional in international political. We further check their micro-blogs records during those three months, and find that those users are relatively high activities, and their micro-blogs are often forwarded, supported by many other users, take 1516940868 as an example, he often went to the areas with intense international conflicts, such as Iraq, Libya, and reported the latest news of current situation on Tencent micro-blogs, his behaviors and opinions are always followed and supported by plenty of users, and lots of users often communicated with him to express their own idea. According to the analysis of experiment results above, TOIM with opinion propagation algorithm could successfully detect opinion leaders and classify them into different domain automatically.

Second, by drawing scatter figures, the distributions of all users' sentiment on a certain topic or objects in micro-blogs network are plotted, which could reflect real economic trend.

*Insert Figure 14 here.*

Figure 14 shows the relation between public opinion and Chinese economic. Specifically, Figure 14 exhibits the positive/negative opinion distribution of all users over the *economics* topic under $O_3$. Obviously, many users are more concerned about development of Chinese economics, although China has achieved great economic success. Such concern corresponds to many serious problems of Chinese economics, like extremely high housing price and larger gap between rich and poor. Figure 13 shows that the changes of all users' positive attitudes toward the topic *finance market* under $O_3$, has a high correlation with China Hushen-300 Index shown in Figure 14. It implies that the public opinion can reflect the real financial situation.

## 6. CONCLUSIONS AND FUTURE WORKS

This paper investigates a novel problem of social opinion influence on different topics in microblog. A Topic-level Opinion Influence Model (TOIM) is proposed using Tencent Micro Blogging, a famous Micro-Blogging website in China, to formalize this problem in a unified framework, and to capture the dual effect of topic preference and social influence on opinion prediction problem. Users' historical messages and social interaction records are leveraged by TOIM to construct their historical opinions and neighbors' opinion influence through a statistical learning process, which can be further utilized to predict users' future opinions towards some specific topics. Gibbs sampling method is introduced to train the model and estimate parameters.

To test and evaluate the proposed model, an experiment was constructed based on a sub-dataset from Tencent Micro Blogging data, and the results show that the proposed TOIM can effectively model social influence and topic simultaneously and clearly outperformed baseline methods for opinion prediction. A distributed TOIM based on the Map-Reduce framework is also proposed to improve the computational efficiency. In conclusion, we demonstrate that the predicted opinion from TOIM can be used for both *microscopic* and *macroscopic* level analysis, including the local opinion influence visualization and global opinion correlation with real-world phenomena.

There are several limitations for TOIM. One main aspect is the misunderstanding of users' true meaning due to the limitation of opinion detection technologies. The reason is that human language expression can be complicate: e,g, mocks, analogy, implication. Even the same word can have different meaning under diverse contexts. Such ambiguity can be further aggravated in microblog environment, where people tend to create short, informal and vague messages. Besides, the opinion detection in our model is still primitive, thus potential challenges include how to design and obtain dedicate topic labels, how to effectively pre-process experiment datasets (e. g., delete noisy information, use training data to define constraint rules for learning algorithms), and so forth.

## 7. ACKNOWLEDGMENTS
This paper is supported by China Post Doc Funding(2012M510027). National Basic Research Program of China (No.2011CB302302). He Gaoji Project, Tencent Company (No.2011ZX-01042-001-002). The National Natural Science Foundation of China(NSFC Program No.71072037).

### Table 1: The Summarization of Tencent Micro-Blogging from 2011 Oct to 2012 Jan

|  | Users | Forwards | Replies | Comments | Mention |
|---|---|---|---|---|---|
| **Total Number** | 326,497,021 | 1,026,243,542 | 43,658,122 | 299,354,146 | 204,345,273 |

### Table 2: Summary of experimental data

|  | # of post message | # of reply message | # of users |
|---|---|---|---|
| Total | 2,350,372 | 959,918 | 145,327 |
| $O_1$ | 320,176 | 114,382 | 24,382 |
| $O_2$ | 591,433 | 243,876 | 31,432 |
| $O_3$ | 742,853 | 298,746 | 38,796 |
| $O_4$ | 472,463 | 275,148 | 28,254 |
| $O_5$ | 295,447 | 136,748 | 22,463 |

### Table 3: Conservative and Non-conservative influence propagation effect on user opinion prediction

| Methods | TOIM | Steps | Conservative | | Non-Conservative | |
|---|---|---|---|---|---|---|
|  |  |  | Average Precision Improved | t-test | Average Precision Improved | t-test |
| Average Precision | 0.5572 | $t=1$ | +0.0021 | >0.05 | -0.00106 | >0.05 |
|  |  | $t=2$ | +0.00235 | >0.05 | -0.00302 | >0.05 |
|  |  | $t=3$ | +0.0009887 | >0.05 | -0.00332 | >0.05 |

### Table 4: Users' opinion preference on different topics

| User ID | User Description | Sentiment on Topics | The most discussed Words |
|---|---|---|---|
| 622007070 | 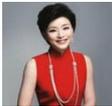 **A famous Host** (Followers Num: 10,116,317) Num of Messages: 28 Num of Response: 30,700 | **Positive:** Topic 1 (*College & Education*): 0.7101<br>Topic 2 (*Daily Emotion*): 0.6845<br>Topic 6 (*Country & History*): 0.4729<br>**Negative:** Topic 1 (*College & Education*): 0.2899<br>Topic 2 (*Daily Emotion*): 0.3155<br>Topic 6 (*Country & History*): 0.5271 | Topic 1: Daughter, Exam, University<br>Topic 2: Happiness, Expectation, Women<br>Topic 6: Women, China, Army, Occupy<br>Topic 1: Primary School, Student<br>Topic 2: Encounter, Home violence, Worry<br>Topic 6: War, Women, Power, Hurt, Pain |
| 394865678 | 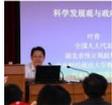 **A famous finance Officers** (Followers Num: 2,186,021) Number of Messages: 36 Number of Response: 6,242 | **Positive:** Topic 1 (*College & Education*): 0.7907<br>Topic 3 (*Social Affairs*): 0.5463<br>Topic 23 (*Country Develop*): 0.4729<br>**Negative:** Topic 1 (*College & Education*): 0.1093<br>Topic 3 (*Social Affairs*): 0.4537<br>Topic 23(*Country Develop*): 0.5271 | Topic 1: Research, University, Student<br>Topic 3: Protection, Children, Driver<br>Topic 23: Innovation, Investment<br>Topic 1: College<br>Topic 3: Accident, Corruption<br>Topic 23: Enterprise, Lawsuit, Stock |
| 1484189007 | 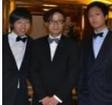 **A famous Intel. Researcher** (Followers Num: 800,779) Number of Messages: 78 Number of Response: 31,644 | **Positive:** Topic 1 (*College & Education*): 0.5777<br>Topic 13 (*Economics*): 0.3227<br>Topic 27 (*Intel. Political*): 0.3939<br>**Negative:** Topic 1 (*College & Education*): 0.4223<br>Topic 13 (*Economics*): 0.6773<br>Topic 27(*Intel. Political*): 0.6061 | Topic 1: University, Student, China<br>Topic 13: Innovation, Medical, Economics<br>Topic 23: Charm, People, Solution, Tough<br>Topic 1: Education, Research, Lost<br>Topic 13: Cost, Technology, Company<br>Topic 23: Relationship, Criticism, Protest |
| 790147482 | 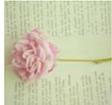 **Vice Editor in Finance Domain** (Followers Num: 146,718) Number of Messages: 47 Number of Response: 14,140 | **Positive:** Topic 13 (*Economics*): 0.2988<br>Topic 14 (*Social Study*): 0.3267<br>Topic 23 (*Country Develop*): 0.5232<br>**Negative:** Topic 13(*Economics*): 0.7012<br>Topic 14 (*Social Studies*): 0.6733<br>Topic 23(*Country Develop*): 0.4368 | Topic 1: Reform, Tax, Finance Developing<br>Topic 13: Democracy, Institution, Law<br>Topic 23: Charm, People, Solution, Tough<br>Topic 1: Industry, Monopoly, Debit<br>Topic 13: Officers, Society, Market<br>Topic 23: Income, Inflation, Welfare |

Table 5: Examples for predicting users' opinions by using TOIM with Conservative Propagations

| User Identification | Methods | Influential Neighbors | | Results |
| --- | --- | --- | --- | --- |
| | | Agree | Disagree | |
| 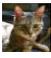Xgdd (Positive) | TOIM with Conservative | 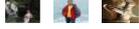 | 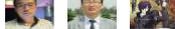 | Correct |
| | CRFs | Null | Null | Wrong |
| 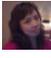Lyh_Lawer(Negative) | TOIM with Conservative | 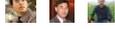 | Null | Correct |
| | CRFs | Null | Null | Wrong |
| 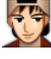Zhang(Negative) | TOIM with Conservative | 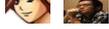 | 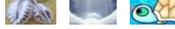 | Correct |
| | CRFs | Null | Null | Wrong |
| 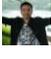Hu_Chunhua(Negative) | TOIM with Conservative | 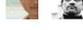 | 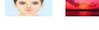 | Correct |
| | CRFs | Null | Null | Wrong |
| 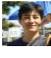Buffaloes (Positive) | TOIM with Conservative | 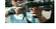 | 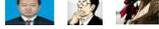 | Correct |
| | CRFs | Null | Null | Wrong |

**Input**: $G = (U, M, A, E)$
**Output**: $\Theta, \Phi, \Psi, \Omega$
Initiation: Iterations $Iter$
Pre1: Generate $W_N$ and $W_O$;
Pre2: Construct parse tree for $m_i \in M$;
Pre3: Calculate $SD(n_i, ow_j), n_i \in W_N, ow_j \in W_O$ based on Equation 1;
Pre4: Calculate $NOAI(u_i, u_j), u_i, u_j \in U$ based on Equation 7;
Pre5: Obtain $CoE$ from Training Dataset;
Start:
**for** $e = 1 : Iter$ **do**
  **for** $m_i$ in $M$ **do**
    find the user $u_i$ who posted $m_i$;
    find all comments $M_i$ on $m_i$;
    **for** each noun $n_i$ in $m_i$ **do**
      sample topic $z_i$ based on Equation 8;
      detect $u_i$'s opinion $o_i$ of $n_i$ based on parse tree or Equation 2;
      **if** $o_i == +1$, $C^{z_i}_{i,+1} += 1$; **else** $C^{z_i}_{i,-1} += 1$;
      **for** each comment or reply $m_j$ in $M_i$ **do**
        find the user $u_j$ who creates $m_j$;
        **for** each noun $n_j$ in $m_j$ **do**
          **if** $n_j == n_i$, set $z_j = z_i$; **else** sample topic $z_j$ based on Equation 8;
          detect $u_j$'s opinion $o_j$ of $n_j$ based on parse tree or Equation 2;
          **If** $o_j == +1$, $C^{z_j}_{j,+1} += 1$; **else** $C^{z_j}_{j,-1} += 1$;
          **if** $z_i == z_j$ **then**
            **If** $(CoE(n_i, n_j) == 1, o_i == o_j)$
              $C^{z_i}_{i,j,agree} += 1$;
            **else if** $(CoE(n_i, n_j) == 1, o_i != o_j)$
              $C^{z_i}_{i,j,disagree} += 1$;
            **else if** $(CoE(n_i, n_j) == 0, o_i == o_j)$
              $C^{z_i}_{i,j,disagree} += 1$;
            **else if** $(CoE(n_i, n_j) == 0, o_i != o_j)$
              $C^{z_i}_{i,j,agree} + 1$;
          **end**
          **if** can not judge opinions between $u_i$ and $u_j$ **then**
            sample $temp$ from $NOAI(u_i, u_j)$;
            **If** $temp <= NOAI(u_i, u_j)$,
            $C^{z_i}_{i,j,agree} += 1$;
          **end**
        **end**
      **end**
    **end**
  **end**
**end**
Calculate $\Theta, \Phi, \Psi, \Omega$ based on Equation 9, 3 and 4.

**Algorithm 1**: Estimation of $\Theta, \Phi, \Psi$ and $\Omega$

```
Input: Ψ, Ω, user $u_j$, topic $t_k$, weight $w$
Output: opinion $o_j^{new}$
Initiation: Iterations $Iter$, $w$
for $e=1:Iter$ do
    randomly sample a user $u_i$ in $ON(u_j, t_k)$;
    $u_i$'s opinion $o_i$ is known or sampled from $\psi_{i,+1}^k$ if unknown;
    set $temp = w\phi_{i,j}^k + (1-w)\omega_{i,j,agree}^k$
    sample $o_j^{temp}$ from $temp$;
    if $o_j^{temp} == o)i$ then
        $SWO+ = o_j^{temp} * s_{i,j,agree}^k$
    else
        $SWO+ = o_j^{temp} * s_{i,j,disagree}^k$
    end
end
if $SWO > 0$ then
    $o_j^{new} = +1$;
else
    $o_j^{new} = -1$;
end
```

**Algorithm 2:** Opinion Prediction

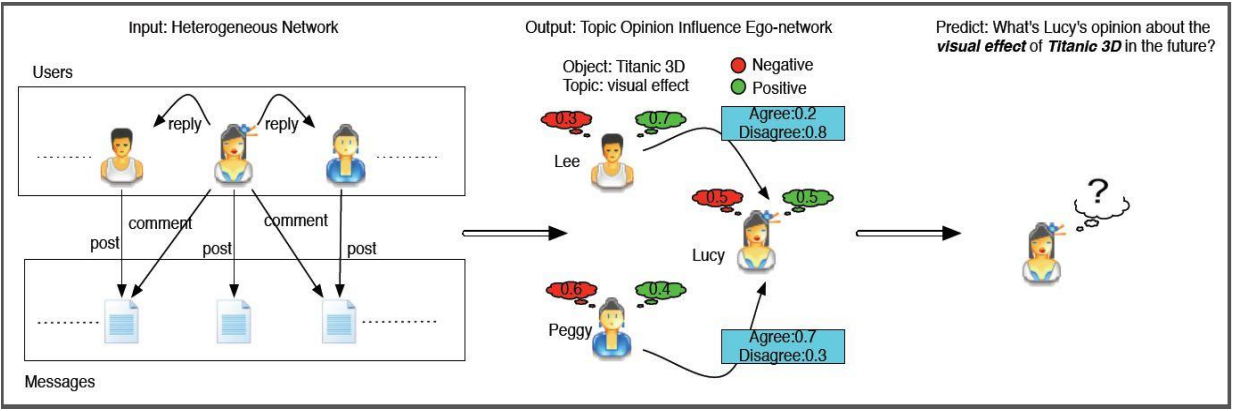

**Fig.1. Motivating Example**

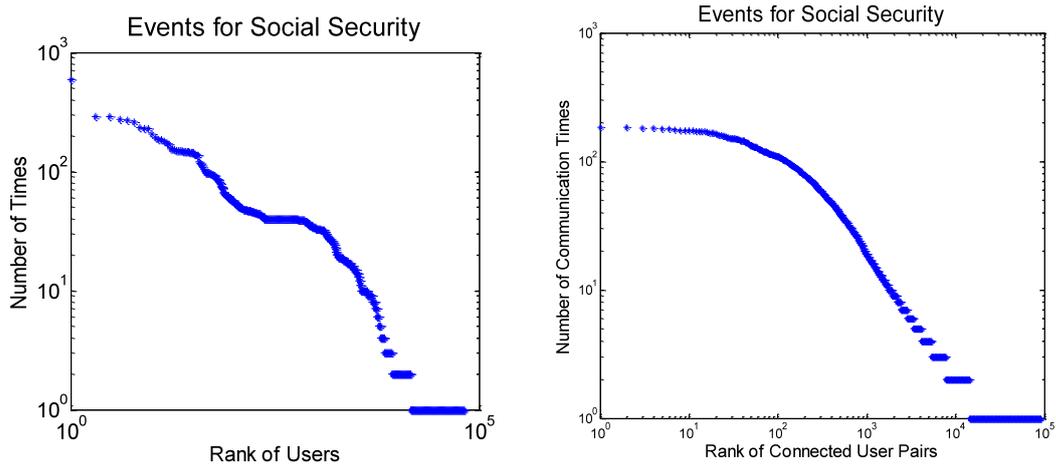

**Fig.2. Statistical Analysis of Users' behaviors**

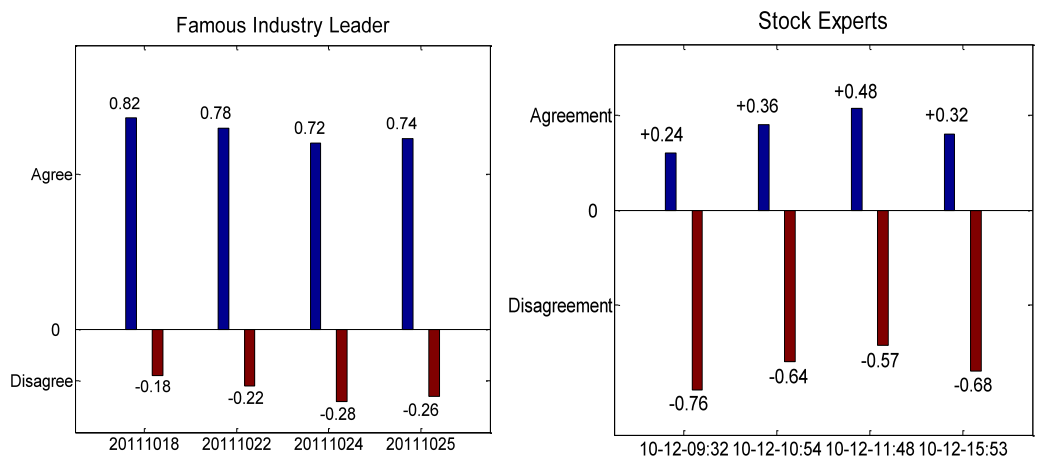

**Fig.3. Two examples of users' opinion influence**

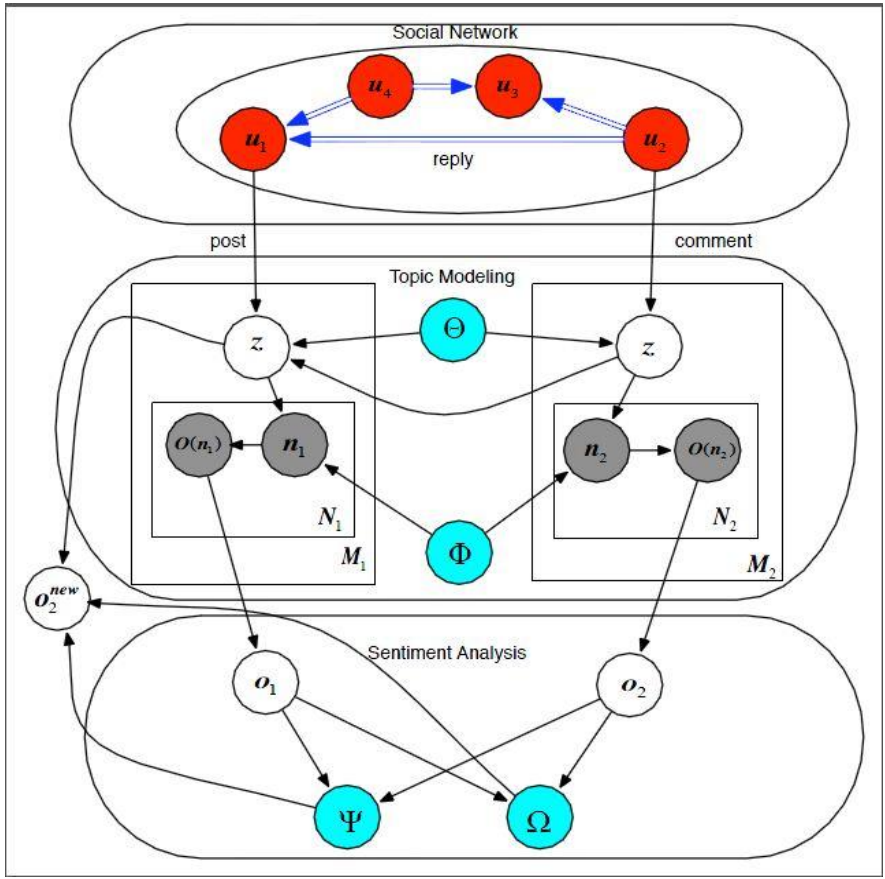

**Fig.4. The Framework of TOIM**

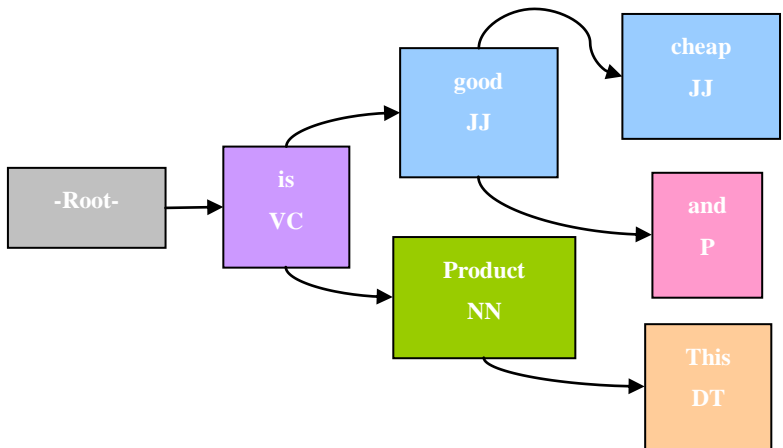

**Fig.5. An example of Parse Tree**

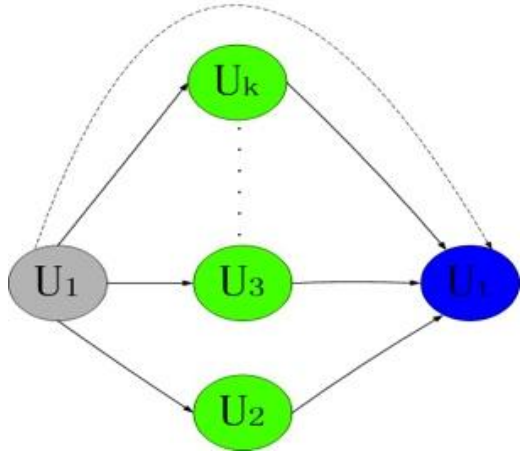

**Fig.6. An example of indirect opinion influence**

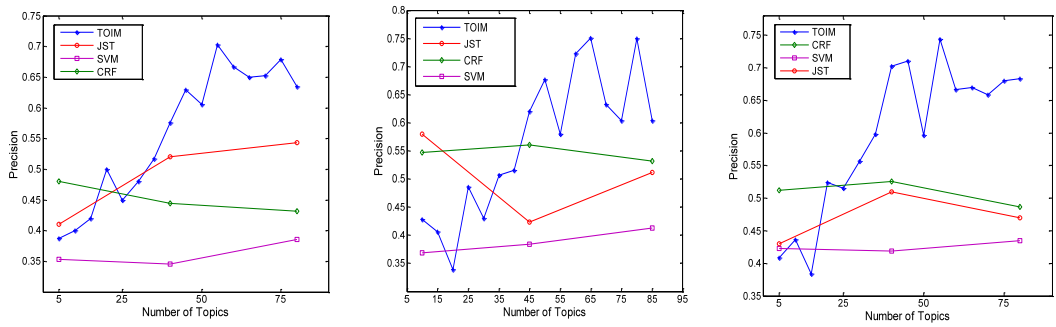

**Figure 7: Opinion Prediction of $O_1$, $O_2$ and $O_3$**

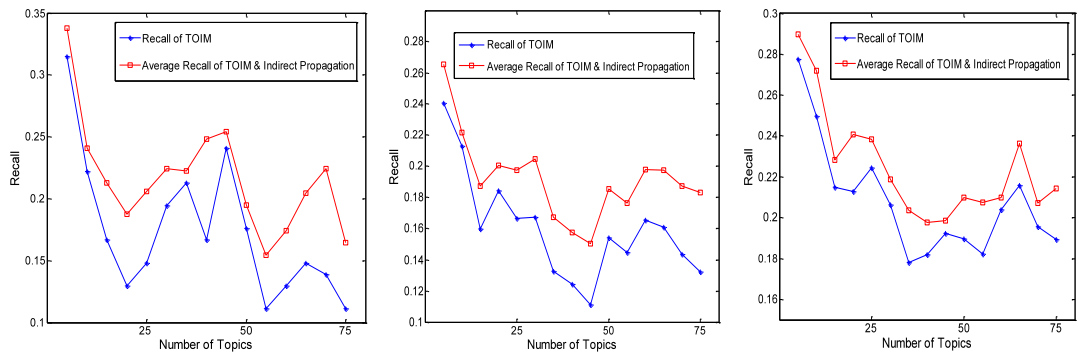

**Figure 8: Recall of $O_1$, $O_2$ and $O_3$**

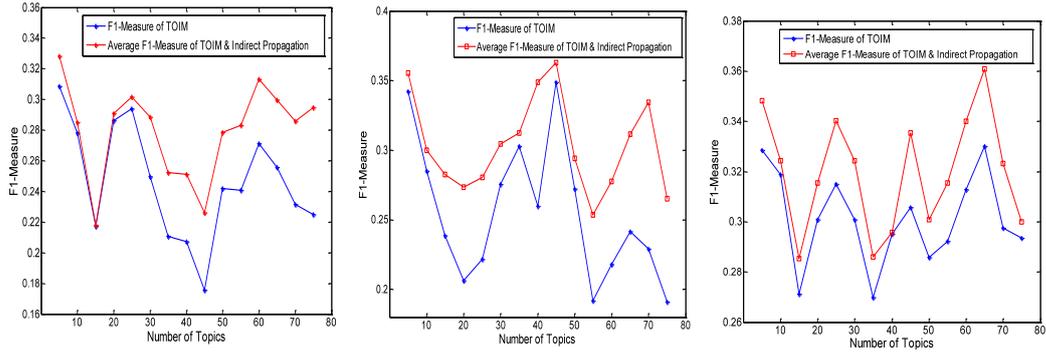

**Fig.9. F1-Measure of** $O_1$, $O_2$ **and** $O_3$

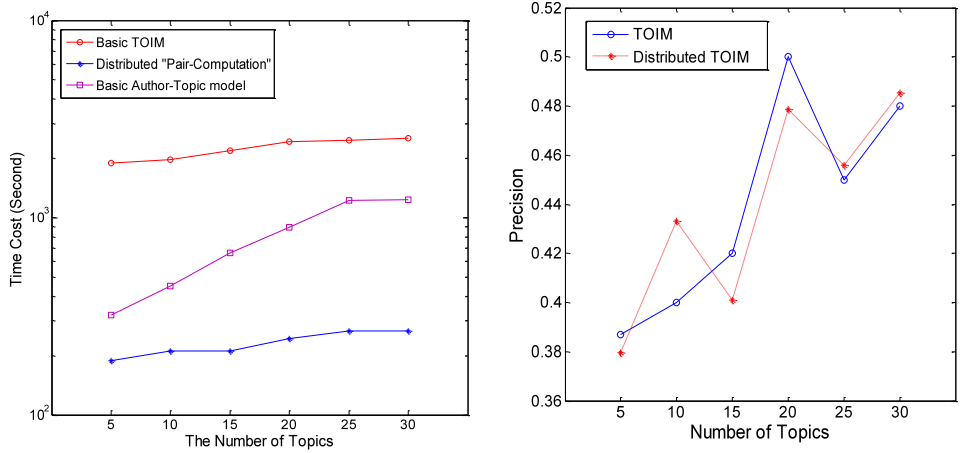

**Fig.10. Time & Precision Comparison of Distributed TOIM and TOIM**

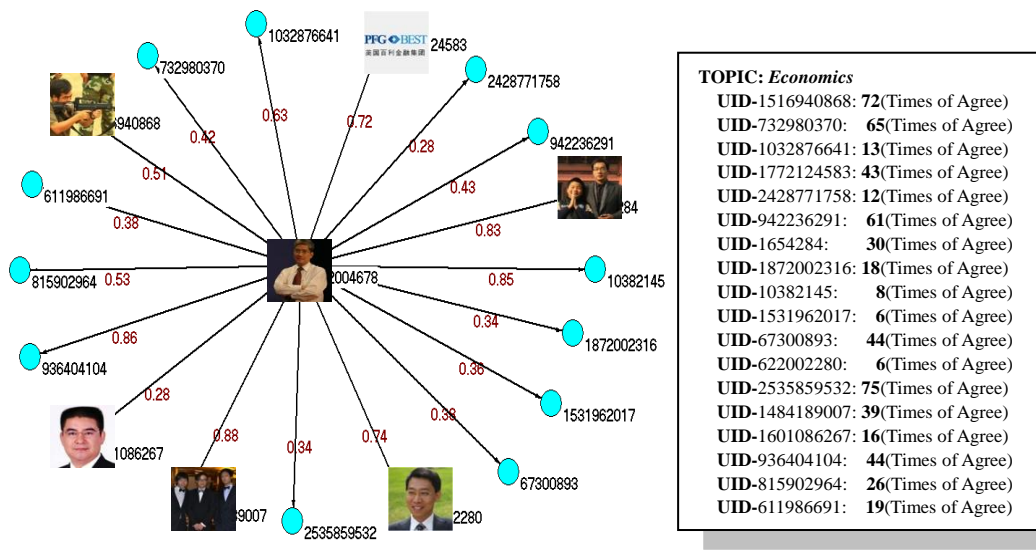

**Fig.11. Examples of Direct Influence By TOIM**

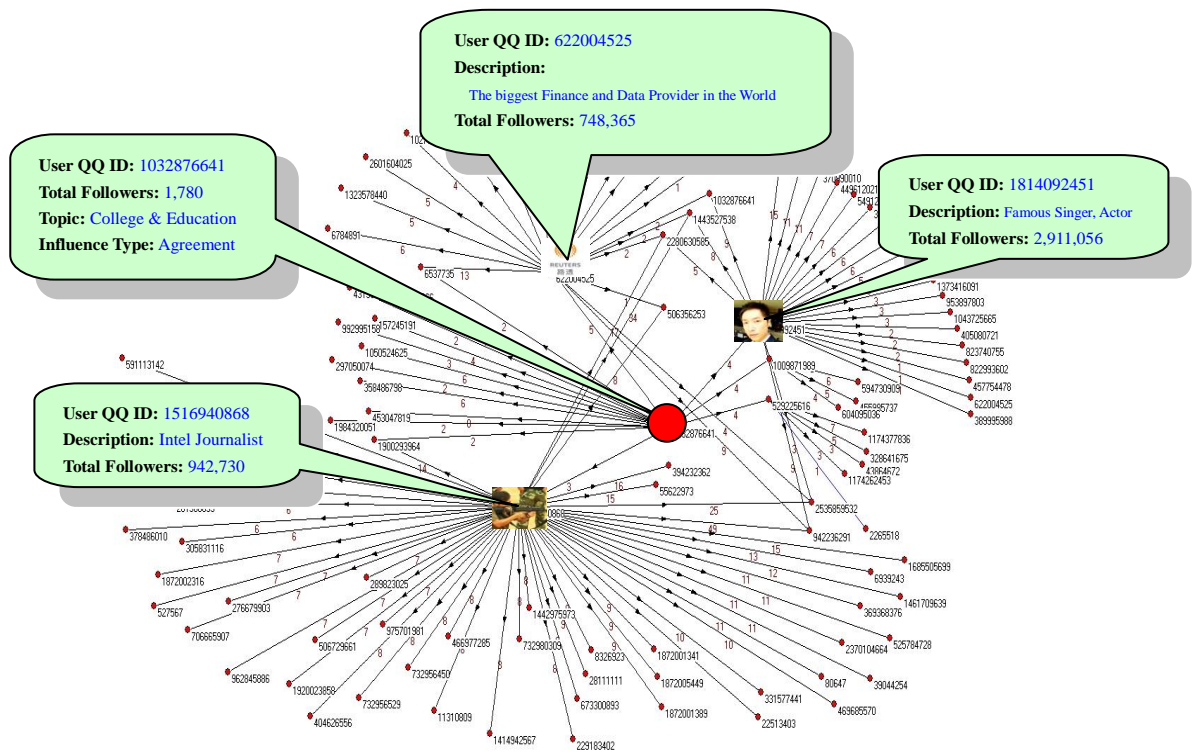

**Fig.12. An Example of Indirect Opinion Influence**

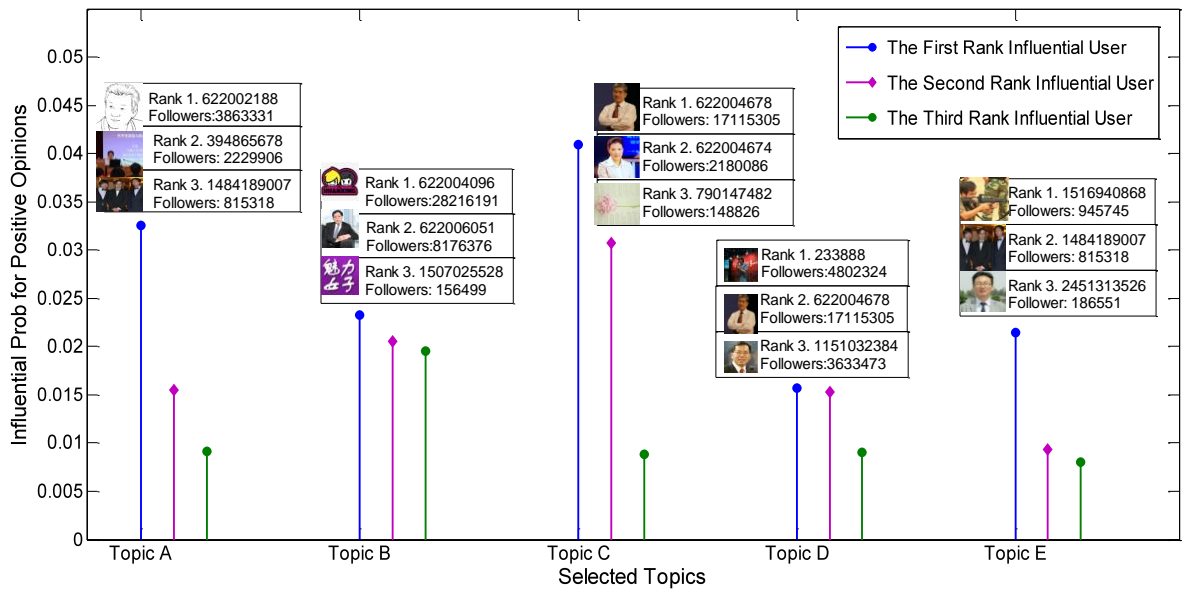

**Fig.13. The most Influential users for different topics and the number of their followers in three months**

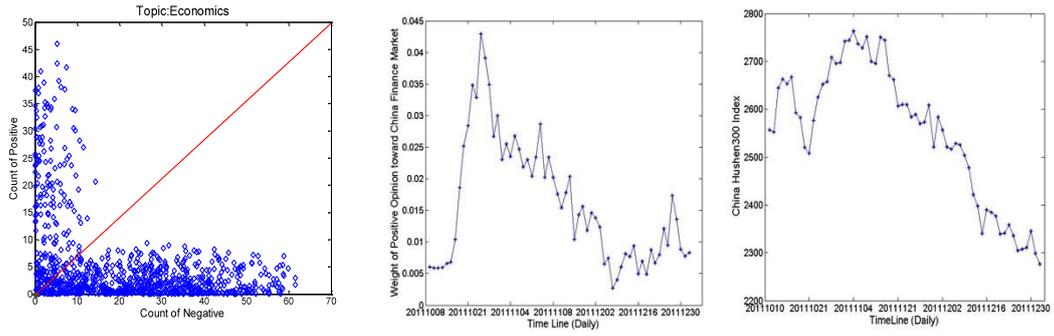

**Fig.14. Correlation between Collective Opinions and Economic Activities**

Response to Editors
[Click here to download Other: Response to Editors_20121009.doc]

Accepted CIKM'12 Paper
Click here to download Other: Accepted CIKM12 Paper.pdf